# Data Accessibility as a Pathway to Genuine Equality for ATLANTA'S WESTSIDE COMMUNITIES

Laura Kathryn O'Connell
Westside Communities Alliance
Atlanta, GA, USA
L.KatieOConnell@gatech.edu

Mackenzie Madden
Westside Communities Alliance
Atlanta, GA, USA
mmadden6@gatech.edu

Sheri Davis-Faulkner, PhD
Westside Communities Alliance
Atlanta, GA, USA
sheridf@iac.gatech.edu

## ABSTRACT

Data is a dominant force during the decision-making process. It can help determine which roads to expand and the optimal location for a grocery store. Data can also be used to influence which schools to open or to shutter and "appropriate" city services to continue or discontinue. Considered fact-based, objective, and impartial, data can trump emotional appeals during the final evaluation of a project; thus creating a power imbalance between those with the resources to access data and those without. Most often left behind are communities already struggling to stay afloat due to years of disinvestment by market forces and external decision-makers. For long ignored residents in Atlanta's Westside neighborhoods, the burden of inaccessible data continuously thwarts their opportunity for mobility. However, with the advent of the internet and the global push for open data, access to information is no longer solely in the hands of those with power, influence and money. Online tools, like the Westside Communities Alliance (WCA) Data Dashboard, quickly disseminate data to those most impacted by "data driven decision-making," thus creating the potential of a genuinely equitable society.

Based out of the Georgia Institute of Technology, the WCA works to build and sustain relationships among constituencies located in West Atlanta with the goal to strengthen partnerships around issues of common concern. The creation of the Data Dashboard stemmed from a recognized community desire for more localized control and the need for improvements to the communities' overall prosperity. Development of the site progressed through significant engagement between the WCA, community groups, and local agencies. The Dashboard takes the vast abundance of data and synthesizes it into a format that is both visually and geographically user-friendly. Through different portals, users can access neighborhood-level data around demographics, housing, education, and history that is formatting in a way that is easily accessible and understandable. The site includes qualitative research that goes beyond data and stats to give a totality of the community. By allowing West Atlanta advocacy groups to easily retrieve data, the WCA Data Dashboard empowers residents, nonprofits, and neighborhood associations to be full participants in the decision-making process.

## 1. INTRODUCTION

In 1967, less than a year before his murder, Martin Luther King Jr. wrote, "It's much easier to integrate a lunch counter than it is to guarantee a livable income and a good solid job. It is much easier to guarantee the right to vote than it is to guarantee the right to live in sanitary, decent housing conditions. It is much easier to integrate a public park than it is to make genuine, quality, integrated education a reality. And so today we are struggling for something which says we demand genuine equality". He recognized that equality was not simply equal protection under the law, but it also meant that all individuals have a chance for equal opportunities. Today, many economic markers show equality further from reality at the national scale. Within some cities, as in the case of Atlanta, the disparities grow further. Data can work to underscore inequality while also giving validation to community advocacy groups.

Neighborhood Information Systems (NIS) disseminate data and allow more organizations to easily participate in the shaping of their future. There are a plethora of NIS around the United States, but few are designed by and for a specific community like the Westside Communities Alliance's (WCA) Data Dashboard. This Dashboard can serve as an example for other communities looking to use data as a tool to help shape their destiny and bring genuine equality to all of their residents.

## 2. EQUALITY

Prior to the 1950s, African Americans were deemed second class citizens throughout the majority of the United States. The South had de jure segregation through Jim Crow that forced blacks to sit in the back of the bus, use different amenities, and attend substandard schools. While the South had systematic policies in place, the North also participated in de facto segregation including redlining - a practice that ensured neighborhoods and schools remained separate and unequal. Starting with the 1955 Supreme Court case, Brown v. Board of Education, there was a push for policies aimed at breaking these systems and working to improve equality. Subsequent legislation like the Civil Rights Act of 1964 and 1968 prohibited discrimination in both employment and housing.

The Civil Rights icon, Martin Luther King, Jr., first championed to ensure the State recognized African Americans as equal citizens, but as anti-discrimination laws were enacted, he turned his attention towards the idea that legal equality was not the same as genuine equality. Simply giving a person access to a broad range of public accommodations does not ensure the same opportunities for employment. The law stated schools must no longer be segregated, yet proportionally African American children continued to go to underfunded and overcrowded schools at significantly higher rates than whites.

At the time, one of the starkest examples of inequality in the United States was in the form of economic inequality. In his 1964 State of the Union address, President Johnson noted "many Americans live on the outskirts of hope -- some because of their poverty, and some because of their color, and all too many because of both". King too recognized the problems of poverty and created the Poor People's Campaign in 1967 with a goal to improve the economic future for all Americans; no matter the race.

Economic inequality among all Americans peaked in 1928, a year before the onset of the Great Depression. At the time, the top 1 percent of income earners received 19.26 percent of the nation's income. Over the next half century, this measure slowly declined until it hit its lowest point at 7.74 percent in 1973. However, a reversal began and over the course of the next decade, the number slowly increased until the mid-1980s when the top 1 percent's income share jumped from 9 to 13 percent. This was the start of the another rapid rise in economic inequality with steep ascension that continued through the end of the Great Recession in 2012 when the top earners' share of income rose to 18.88 percent nearly the same as 1928 (Saez 2014). Described a different way, in 2012 the top 10 percent of earners brought in 9 times as much income as the bottom 90 percent, while the top 1 percent had 38 times more income. Moreover, the nation's top 0.1 percent earned 184 times the income of the bottom 90 percent; which meant a full-time minimum wage worker would have to work 930 years to earn as much as a fast food CEO earned in a single year (Nichols 2014).

Today, across the United States, economic inequality continues to rise with Atlanta having the greatest disparities. Looking at the ratio of top 95th percentile earners to bottom 20th percentile earners, Atlanta has the greatest ratio at 19.2 (ACS 2007-2013). The upper limit for bottom earners was $14,988 and the lower limit for top earners was $288,159. The next highest cities are San Francisco at 17.1 and Boston at 15. This situates Atlanta at nearly twice the national average of 9.3 (Berube 2014).

Within the City of Atlanta, some of the highest inequities are in West Atlanta, which is an area just west of the top tourist destinations in the city and has a population that is 88 percent African American with some census tracts as high as 97 percent. Approximately 54 percent of households have an income below $50,000 compared to 30 percent of the City of Atlanta, while the overall unemployment rate is 10 percent compared to Atlanta's 8 percent. Sixty-two percent of children under five live in poverty in West Atlanta almost doubling Atlanta's 38 percent. It is recognized that households should not spend more than 30 percent of their income on housing costs, yet in West Atlanta over 43 percent of the population spends more than 50 percent of their income on gross rent compared to 29 percent of Atlanta. Finally, the housing vacancy rate is 36 percent compared to 20 percent with much of the stock abandoned and uninhabitable.

## 2.1. DATA FOR GENUINE EQUALITY

To push against this growing inequality, it is imperative that local non-profit organizations, neighborhood advocacy groups and the residents ("the Community") act as guiding agents for the residents to combat the change happening in their communities. Residents are often left out of critical conversations about their neighborhoods' future because they are considered not knowledgeable in the technical details that governmental agencies ("agencies") require. Community organizations are left with only emotional appeals that are often not supported by facts. As agencies are further pushed to make evidence-based decisions, the role of data in the decision making process takes greater importance and the Community is shut out. By breaking down the barrier to data, the Community can have more opportunity to make the case for governmental decisions that will ultimately result in parity in services and amenities at the neighborhood level that offers the potential for greater equality.

Data accessibility works on a number of fronts in achieving the goal of genuine equality. Data helps the Community keep a watchful eye on agencies to ensure a timely follow through on promises, leading to increased agency transparency. The development of trackable metrics improves accountability by ensuring agencies stay true to the goal and promises. It improves collaboration with the Community not just in planning, but in implementation as it is easier to work together when everyone has equal access to information. It reduces duplicating efforts. Data can increase citizen participation by making individuals feel more informed by participating in the process, building confidence to understand what is presented and make informed decisions. Finally, data can improve efficiency because the Community is no longer envisioning optimal solutions from a historical and emotional vantage point but instead can rely on analysis based on the facts of the situation.

Data is clearly an important tool for the Community, but the real barrier is accessing and analyzing, making the data usable. For some, the staff does not possess the technical knowledge, while others only focus their energy on advancing their cause while not doing time-consuming data crunching and mapping. A solution to this problem is through Neighborhood Information Systems ("NIS"), an online tool that includes data and geography. The tools are often designed by third parties and are designed with the Community in mind.

As technology has improved over the last decade, NIS have been rapidly appearing. Some display data, others are created as indicator tools, while others are calculators. Each of these are useful for their abilities, but many are designed for advanced computer users or people with advanced data knowledge. Another problem is the overwhelming number of sites that users must keep track of and the fact that different sources may be used for the same indicators, producing different results. Finally, a number of sites are not designed at a meaningful geographic scale. Either sites are hyper localized at the census tract level with little meaning to the lay person or they are scaled up to county level. For some localities, county level geography can blur the facts. This is true in Fulton County where Atlanta is located which is characterized by demographic data that swings widely from north the south. For organizations working at a neighborhood scale, county data does not provide a finite enough description of the community. In Atlanta for example, the Neighborhood Planning Units, the recognized source for citizen participation and governance, is comprised of multiple neighborhoods and census tracts, organized around smaller neighborhood organizations and meeting on a monthly basis. Residents of the NPU who attend have a chance to hear information from local agencies like the police department or code enforcement. They vote on zoning changes and business licenses and have an opportunity to hear about development or plans that will impact their community. However, they are rarely presented with data that is specific to their NPU that could lead to the most informed decision making. For many Atlanta advocacy groups, data at the NPU level is most desirable.

# 3. WCA DATA DASHBOARD

The Westside Communities Alliance is a communications network sponsored and managed by the Ivan Allen College ("IAC") of Liberal Arts at Georgia Tech. Created in 2011 in response to the problems faced by West Atlanta ("Westside"), the WCA "works to build and sustain relationships among other constituencies across communities west of the Connector, and with others, in order to partner on issues of common concern". The WCA defines Westside as the communities located within NPUs K, L, and T. This area lies adjacent to the western edge of Georgia Tech and encompasses the communities of English Avenue, Vine City, as well as the Atlanta University Center Neighborhood ("AUCN") that contains the largest number of contiguous Historically Black Colleges and Universities in the United States including Clark Atlanta University, the Interdenominational Theological Center, Morris Brown College, Morehouse College, Morehouse School of Medicine, and Spelman College ("the AUC").

Since its inception, the WCA recognized the importance of data dissemination as a tool for community development. Dean Royster of IAC believed a NIS, in the form of a Data Dashboard (the "Dashboard"), could work to improve local conditions, and in 2014 the WCA began formal development of an online tool. The WCA Data Dashboard officially launched in 2016.

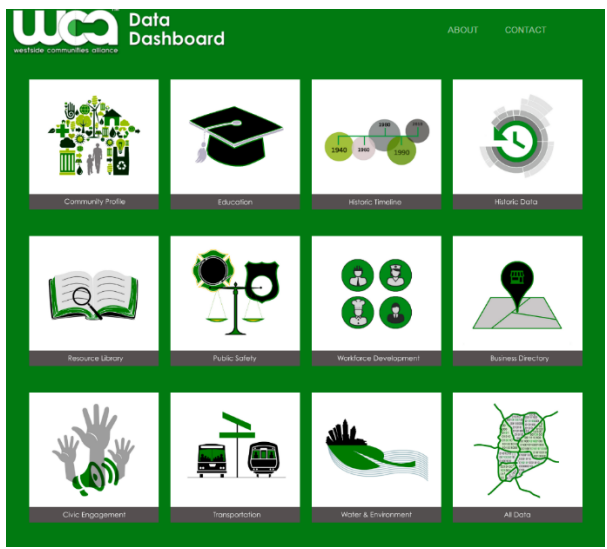

**Figure 1 - WCA Data Dashboard**

The Dashboard is a one-stop data shop presented in locally recognized and meaningful geographies. Most of the data available on the Dashboard includes neighborhoods throughout the city; this is done primarily for comparison purposes. The site design focuses on the Westside, and portal development draws from their requests. Currently there are six main portals with several more in the pipeline - community profile, education, historic data, historic timeline, resource library, and public safety.

The community profile page contains primarily census data demonstrated at three different geographies - NPU, neighborhood, and school cluster. General topics include demographics, housing, poverty, and income. A user can navigate to their desired geography and easily switch between different datasets. Data is presented as bar charts and each variable is compared against the City of Atlanta.

The education portal is very similar to the community profile page, but focuses on school specific data such as enrollment, standardized test scores, and Atlanta Public Schools' enrollment. Users can look at the aforementioned datasets by general student body as well as subcategories like race, gender, economic standing, and disability status. The portal allows users to easily compare across school clusters.

Looking at current data is important for organizations working on grant funding or making the case for improved public services, but it is also important to understand the community from a historic context. Both the historic data page and timeline address this need. The historic data page displays NPU-level decennial census data of race, homeownership, and housing tenure from 1940 to 2010. The user can view data as a bubble, bar, or line graph, which is then displayed using an animated motion chart. The use of movement is especially impactful when watching rapid changes in population, which are not as dramatic when viewing a static graph.

It is important to recognize that communities are not just a series of data points, but rather a dynamic population with historical roots. This is especially true of the Westside, which was the center of the Civil Rights' student movement and includes the former home of many Civil Rights leaders including Martin Luther King Jr.. The historic timeline gives context to the data by presenting events and policies (both local and national) that possibly affected Westside residents. Events are organized into categories (ex. politics, education) and displayed as an interactive timeline where users can read about the event, access the original source, and hear available audio/visual from the event. Users can view events either chronologically or in category bands. By viewing in bands, the user can view which events happened simultaneously.

Over the past half century, there have been countless community plans developed by Agencies for the neighborhoods of the Westside with few visible outcomes. Similarly, local universities use the residents of the Westside as test subjects with few recommendations to remedy concerns. Ultimately, the community feels over studied, with long festering unresolved issues and the same problems existing decade after decade. The purpose of the resource library component of the Dashboard is to counter this issue by presenting articles, studies, and plans for residents and community groups to quickly access. They no longer have to spend time searching for information pertaining to their community.

The newest portal is the Public Safety page and was developed in partnership with Georgia Tech's Atlanta Data Science for Social Good program. The portal includes crime data, code violations, and community assets displayed using both maps and graphs. The maps allow organizations to pinpoint crime hotspots or determine the built environment's impact on crime. The graphing function demonstrates changes over time and helps organizations determine the optimal timeframe to increase or decrease crime prevention programs.

## 3.1. PROCESS

The WCA describes the broad range of skills and knowledge of the Dashboard audience as "PhDs to Grandmas". This is a challenge as the site must be navigable by users with low digital literacy while also being useful for researchers interested in deeper dives into demographic data. To achieve this, the WCA relies on easy navigation and simple graphics with the goal that users can find their answer in no more than 4 clicks.

The WCA dashboard differs from many NIS because it was based on a continuous dialogue between the site developers at the WCA, other local universities, the Community, and Agencies. While site construction took place at Georgia Tech, most of the Dashboard development evolved from community requests and suggestions. This took place at formal presentations, structured focus groups, and one-on-one meetings over a period of eighteen months.

Prior to the development of the Dashboard, the WCA relied on a static asset map that highlighted important projects and sites in Westside Atlanta. This document demonstrated an area that was more than abandoned businesses and vacant houses, comprised of many assets and not obvious deficiencies. Visibility to the numerous historic structures, the active community organizations and committed and capable residents created a vibrancy that was rarely discussed on the evening news.

Dean Royster recognized that an online tool was needed to disseminate data as well as to reshape the narrative about Westside communities. She envisioned a Data Dashboard designed by and for organizations working on the ground. The earliest digital presence for the Dashboard was created by a Ga Tech graduate-level Information Visualization class. The site allowed users to overlay public transit information, walkability, and demographic data heat map. Data was displayed at the census tract level.

Once the transportation site was online, the WCA hosted a community educational event focused on teaching organizations about ways to utilize online data portals. Attendees ranged in age from mid-20s to late-60s and possessed a diversity of computer skills. Participants walked through local and national sites as well as the current version of the Dashboard and examined the usability of each one. They also discussed what data needs were not being met.

Comments from this session helped the WCA continue to develop the Dashboard. One of the biggest takeaways was the overabundance of choices. Participants were introduced to ten sites each with different information and it was clear that individuals felt burdened with keeping track of all the options. This created the idea that the Dashboard should be a 1-stop shop for data and information pertaining to the Westside communities. This took the Dashboard from a single transportation focused page to a site with seven main topics – community profile, economic development, transportation, education, public safety, and historic information.

For some participants, the demonstrated sites were too complicated as the sites were designed with advanced user interfaces. The neighborhood users wanted a site that presented most of the information up front and did not require them to self-select from a long list of variables or a variety of graphics. Many of these same users found sites hard to read and decipher how and where to find information. To follow the desire to streamline data accessibility, the site is designed with portals presented up-front. Users are not required to wade through information to find their topic area of interest. Instead they are able to click on their topic portal which leads directly into the information. Once on individual portals, the user rarely has to click through multiple variables. Rather it is designed so that the data is already analyzed and presented in a simplistic form. No extraneous information or steps are required.

Two of the comparison Atlanta focused sites included limited data at the NPU level. All of the participants wanted to see more information presented at locally recognized geographies. Census tract data was of little use for them. For education-focused organizations, demographic data described by school zone was very important yet rarely done. The WCA made differing geographic scales a priority in development of the Dashboard. Appendix 1 displays the data structure of the site.

Taking all of the suggestions, the WCA began to work on a Dashboard prototype. Once the site was fully functioning, the WCA did a series of portal themed focus groups with invites to individuals with associated subject matter expertise. Attendees included community groups, governmental agencies, neighborhood associations, and university professors. Working with individuals with detailed knowledge allowed for a deep dive analysis of each portal.

For many of the portals, participants asked for an expansion of the included data. For example, the community profile page included only basic demographic data. From this focus group recommendation, it has expanded to fifty data points. Having field experts allowed the WCA to ensure data presentations made sense. For example, the education data was presented by race. Experts pointed out that most Atlanta Public Schools were nearly 100 percent African American so it was unnecessary to display by race.

From the suggestions of the focus group, the WCA updated the site. Some portals saw minor changes while others, like education, were completely redesigned. Figure 2 describes the Dashboard's timeline from inception through launch.

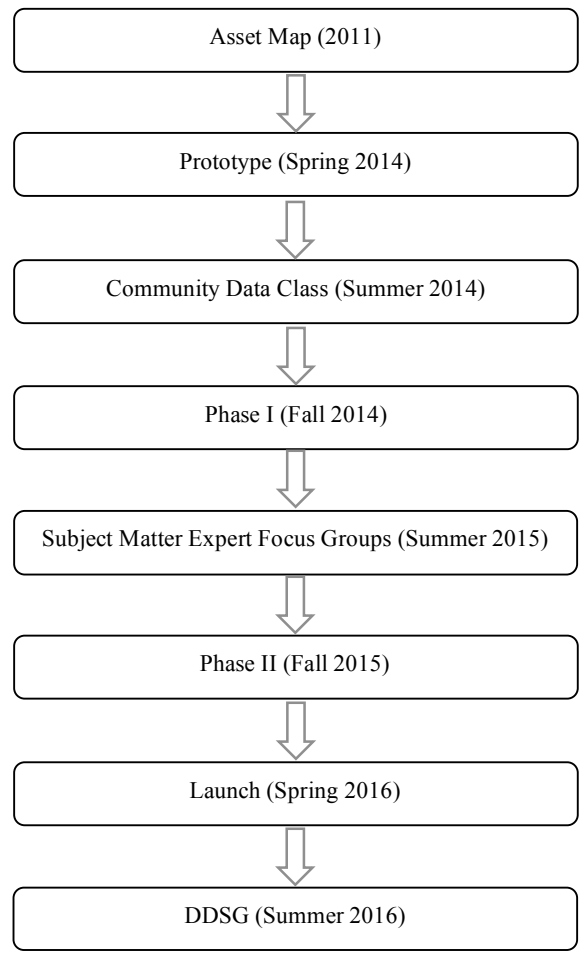

Figure 2 - Dashboard Timeline

4. C

The Dashboard works across a variety of uses and organizations. Local agencies employ the site to do an extended analysis of their community. Nonprofits can access pertinent demographic data for grant funding. An educational advocacy group came prepared with Dashboard data to a critical meeting with APS' Superintendent. It not only increased the credibility of the advocacy group, but also helped shift the conversation in favor of their proposals.

Just as a community is constantly evolving, so too is the WCA Data Dashboard. There are four main projects planned for the next few months. The community profile page will see an expansion of geographies to also include police zones and City Council districts. The data for the community profile and education portals will include an increase in the number of years available. The public safety portal will be demonstrated to a focus group including police, religious leaders, NPU public safety chairs, juvenile justice activist, and the Department of Justice. Comments from this event will shape edits to the portal with a planned public launch by the end of 2016. A number of community organizations have requested the development of an environment and water portal. The WCA will work with these organizations to outline and design an optimal tool. Finally, there will be an increase in qualitative research in the form of storytelling to be added to the historic timeline.

Equality is more than having a right to vote or access to the same public spaces; it is having equal access to high quality schools, meaningful job opportunities, and high quality housing. Data access does not guarantee equity for individuals or between neighborhoods and areas of the city, but can act as a tool to help the Community to participate in decision making and to make visible the inequities of their situation to the general public.. Building a community-driven NIS is the fastest way to disseminate important data sets in an easy to read manner. The Dashboard not only brings data to the Community but changes the narrative by highlighting the assets rather than the more visible deficiencies. Used properly, the power structure for decision making can shift from the Agencies to the Community with visible data that requires accountability and transparency. The WCA's Data Dashboard can serve as a model for other communities in the US interested in achieving greater equity and promoting genuine equality.

## 5. ACKNOWLEDGEMENTS

We acknowledge the ongoing partnership and support of Georgia Tech's Ivan Allen College of Liberal Arts, the College of Design, and the Office of Government and Community Relations as well as technical guidance by members of the College of Computing. We also want to recognize the tireless work of our Westside community partners and residents.

# 7. APPENDIX 1

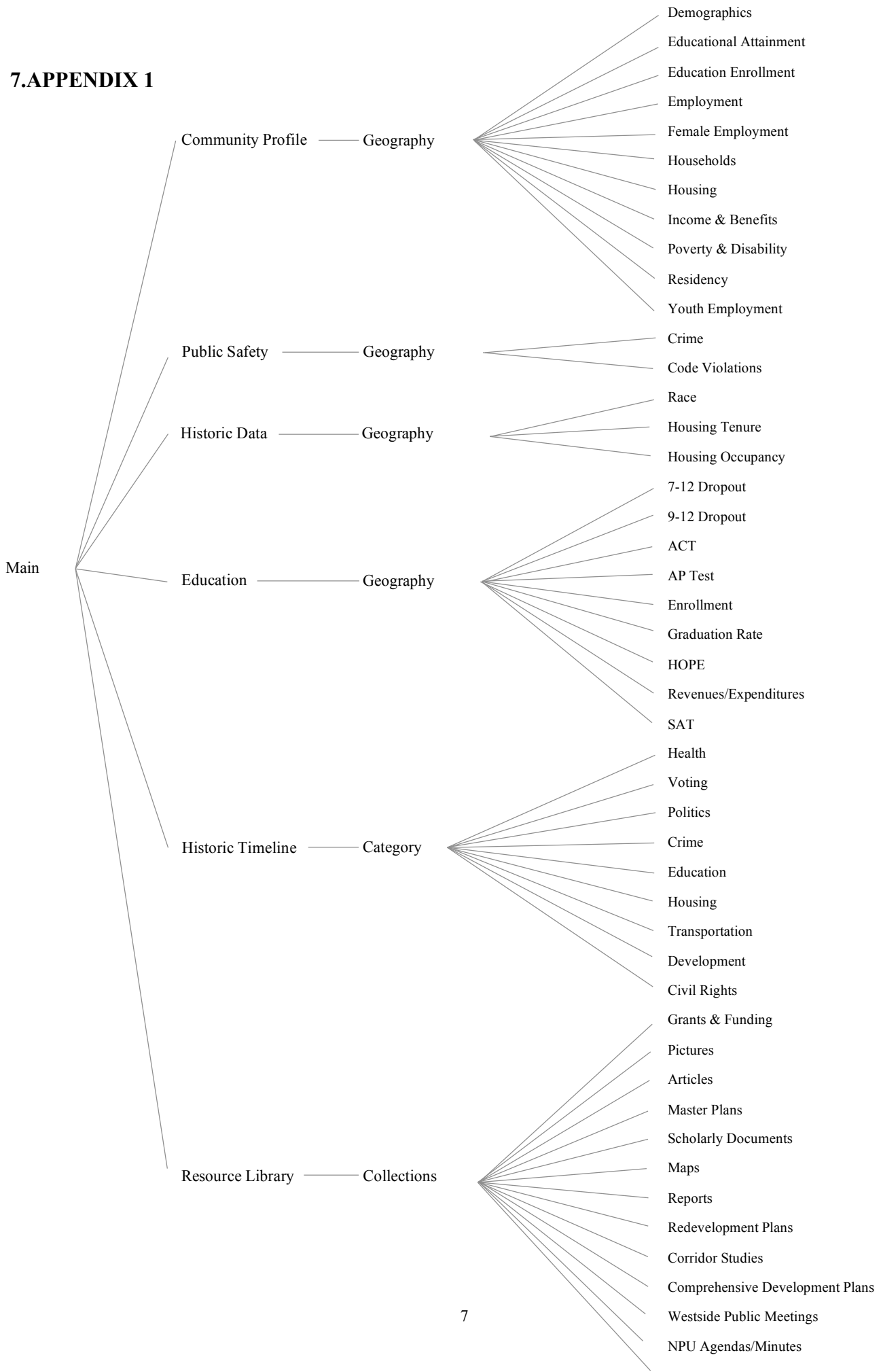